\documentclass[
    ,final            
  ]
  {aipproc}

\layoutstyle{6x9}

\newcommand{\be}{\begin{equation}}
\newcommand{\ee}{\end{equation}}
\newcommand{\bea}{\begin{eqnarray}}
\newcommand{\eea}{\end{eqnarray}}

\def\lsim{\mathrel{\rlap{\lower4pt\hbox{\hskip1pt$\sim$}}\raise1pt\hbox{$<$}}}
\def\gsim{\mathrel{\rlap{\lower4pt\hbox{\hskip1pt$\sim$}}\raise1pt\hbox{$>$}}}
\def\nostrocostruttino#1\over#2{\mathrel{\mathop{\kern 0pt \rlap
{\hbox{$#1$}}} \hbox{\kern-.135em $#2$}}}

\def\kt{k_\perp}

\def\pp{p_\perp}

\def\pt{P_T}

\def\T{_{_T}}

\def\xb{x}
\newcommand{\zh}{z}
\begin{document}
\title{Collins Effect from Polarized SIDIS and $e^+e^-$ Data}


\author{\underline{A.~Prokudin\footnote{Presented by A. Prokudin, e-mail: prokudin@to.infn.it}\hskip 0.1cm}}{
  address={Dipartimento di Fisica Teorica, Universit\`a di Torino and \\
          INFN, Sezione di Torino, Via P. Giuria 1, I-10125 Torino, Italy}
}

\author{C.~T\"{u}rk}{
  address={Dipartimento di Fisica Teorica, Universit\`a di Torino and \\
          INFN, Sezione di Torino, Via P. Giuria 1, I-10125 Torino, Italy}
}

\begin{abstract}
\noindent
The recent data on the transverse single spin asymmetry $A_{UT}^{\sin(\phi_h+\phi_S)}$
from HERMES and COMPASS Collaborations are analysed within LO parton model with 
unintegrated parton distribution and fragmentation functions. A fit of SIDIS data from HERMES Collaboration is performed leading to the extraction of favoured and unfavoured Collins fragmentation functions. A very good description of COMPASS data is obtained. BELLE $e^+e^-$ data are shown to be compatible with our estimates based on the extracted Collins fragmentation functions.
Predictions for $A_{UT}^{\sin(\phi_h+\phi_S)}$
asymmetries at JLab and COMPASS operating on a proton target are given. 
\end{abstract}

\maketitle

\paragraph{\bf \label{Intro} Introduction}

The transversity distribution function, usually denoted as 
$h_1(\xb)$ or $\Delta_T q(\xb)$, is one of the three
fundamental distributions of partons inside a nucleon. It represents 
the distribution of transversely polarized quarks in a transversely polarized nucleon. $\Delta_T q(\xb)$ is so far unmeasured, and it is of great importance to gather all possible information both from the experimental and theoretical points of view. 

Recent data from  HERMES \cite{Airapetian:2004tw} and 
COMPASS \cite{Alexakhin:2005iw} Collaborations on the Collins effect 
\cite{Collins:1992kk} in SIDIS 
gives a strong evidence that the convolution of the Collins 
fragmentation function, that describes the fragmentation of
a transversely polarized quark into an unpolarized hadron (denoted as $\Delta ^N D_{h/q^\uparrow}(\zh, p_\perp)$ or $H_1^\perp(\zh, p_\perp)$, see Ref. \cite{Bacchetta:2004jz}), and the 
transversity distribution function is not vanishing.
Moreover, the data from BELLE \cite{Abe:2005zx} Collaboration on 
azimuthal asymmetries in $e^+e^-$ annihilation reveals that the Collins 
function itself is not vanishing. 

In this work we perform a fit of HERMES \cite{Airapetian:2004tw} data on the SIDIS single spin 
asymmetry  $A_{UT}^{\sin(\phi_h+\phi_S)}$  in order to extract the Collins 
fragmentation function. The validity of the model will be verified by comparing the results of our calculations with 
COMPASS \cite{Alexakhin:2005iw} data. The combined description of BELLE data 
\cite{Abe:2005zx} in terms of the convolution of two Collins fragmentation functions will provide some hints on the size of  the transversity distribution function.
  
We will then give predictions for the azimuthal asymmetries measured at forthcoming JLAB and CERN experiments.

\paragraph{\bf \label{Formalism} Fit and Results}

Spin effects in SIDIS are closely connected to the non-perturbative dynamics 
of the nucleon, and in this contest  the inclusion of intrinsic 
$\kt$ of partons with respect to the nucleon light cone direction plays a crucial role.
Therefore we deal with distribution and fragmentation functions, 
$\Delta_T q(\xb, \kt)$ and 
$\Delta^N D_{h/q^\uparrow}(\zh, p_\perp)$, which explicitely depend on 
the transverse intrinsic momenta of the quarks with respect to the parent 
proton, $\kt$, and of the hadron  with respect to the light cone direction of the 
fragmenting quark, $\pp$.

Leading twist expressions for both the Collins effect in SIDIS
\cite{Collins:1992kk, Kotzinian:1994dv, Mulders:1995dh} and in
$e^+e^-$ annihilation \cite{Boer:1997mf} are well known.
The convolution of the transversity function with the Collins function, 
$\Delta_T q(\xb, \kt) \otimes \Delta^N D_{h/q^\uparrow}(\zh, p_\perp)$,
generates the observed $\sin(\phi_h+\phi_S)$ dependence in SIDIS, whereas the convolution of quark 
and antiquark Collins fragmentation functions, 
$\Delta^N D_{h/q^\uparrow}(z_1, p_\perp) \otimes 
\Delta^N D_{h/\bar q^\uparrow}(z_2, p_\perp)$, gives origin to the 
$\cos( 2\phi_{h_1})$ azimuthal asymmetry in $e^+e^-$ collisions 
\cite{Boer:1997mf}.

For the distribution and fragmentation functions
we will assume a factorized form with a $\kt$ ($\pp$) Gaussian 
dependence, which is very convenient for the description of non-perturbative 
effects at small $\pt$.
Taking also into account the proper behaviour for 
$\kt \to 0$ and  $\pp \to 0$, we have:
\bea
\Delta_T q(\xb, \kt) &=& 
\frac{e^{-{\kt^2}/{\langle\kt^2\rangle\T}}}{\pi \langle\kt^2\rangle\T}\;
\Delta_T q(\xb)\;,  \\
\Delta^N D_{h/q^\uparrow}(\zh, p_\perp) &=& 
\sqrt{2e} \frac{p_\perp}{M_{h}}\, e^{-{p_\perp^2}/{M_{h}^2}} \,
\frac{e^{-{p_\perp^2}/{\langle p_\perp^2\rangle}}}
{\pi \langle p_\perp^2\rangle}
\Delta^N D_{h/q^\uparrow}(\zh)
\;.
\label{collins}
\eea
At this preliminary stage we fix the $\xb$ dependence of the transversity distribution function
by saturating the Soffer bound \cite{Soffer:1994ww}. The $z$ 
dependence of the Collins function is modelled in terms of the unpolarized 
fragmentation function $D_{h/q}(\zh, Q^2)$, as follows
\bea
&& 
\Delta_T q(\xb) = \frac{q(\xb) + \Delta q(\xb)}{2}\;,
\\ 
&&
\Delta^N D_{h/q^\uparrow}(\zh) = 
2 N_q \;
\zh^{\alpha_q} (1-\zh)^{\beta_q} (\alpha_q + \beta_q)^{(\alpha_q + \beta_q)}/({\alpha_q}^{\alpha_q} {\beta_q}^{\beta_q})\; D_{h/q}(\zh)\;,
\eea
with $-1\le N_q\le 1$  to  ensure that the positivity condition
\be
|\Delta^N D_{h/q^\uparrow}(\zh, p_\perp)| \le 
2 D_{h/q}(\zh, p_\perp)
\ee
is respected. Note that we will take positive $\Delta_T u$ and  
negative $\Delta_T d$. We use $q(\xb)$ and $\Delta q(\xb)$ from Refs.~\cite{Gluck:1998xa,Gluck:2000dy}.

The $Q^2$-evolution of transversity is different from that of the unpolarized and helicity distributions.
Nevertheless, since we mainly deal with 
large $x$ and low $Q^2$ values, this difference is negligible. 
The evolution of the Collins fragmentation function, still unknown, is set to be the same as 
that of the unpolarized fragmentation function.

\begin{figure}[t]
\includegraphics[width=0.35\textwidth,bb= 10 140 540 660,angle=-90]
{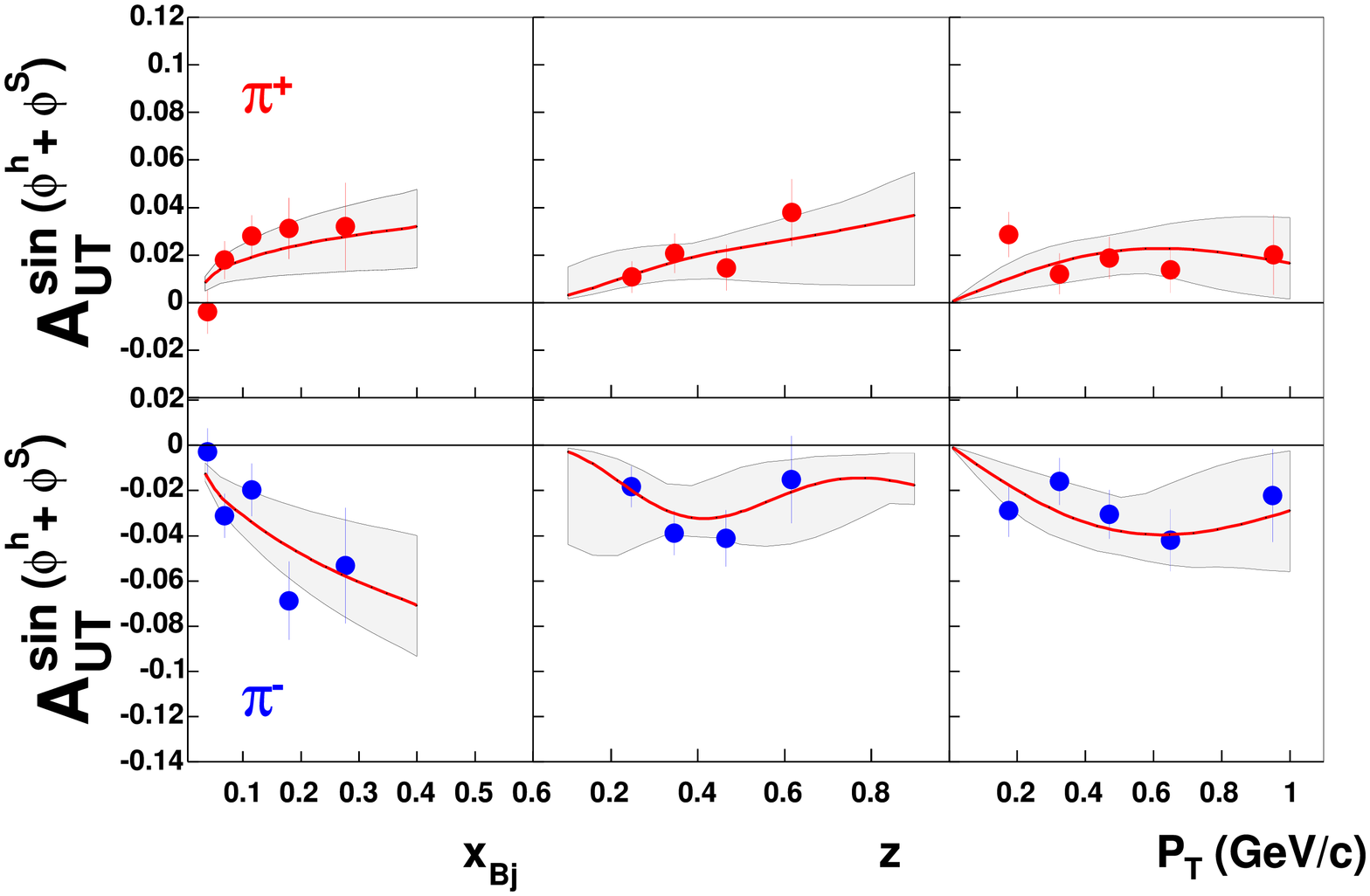} \hskip 1.85cm
\includegraphics[width=0.35\textwidth,bb= 10 140 540 660,angle=-90]
{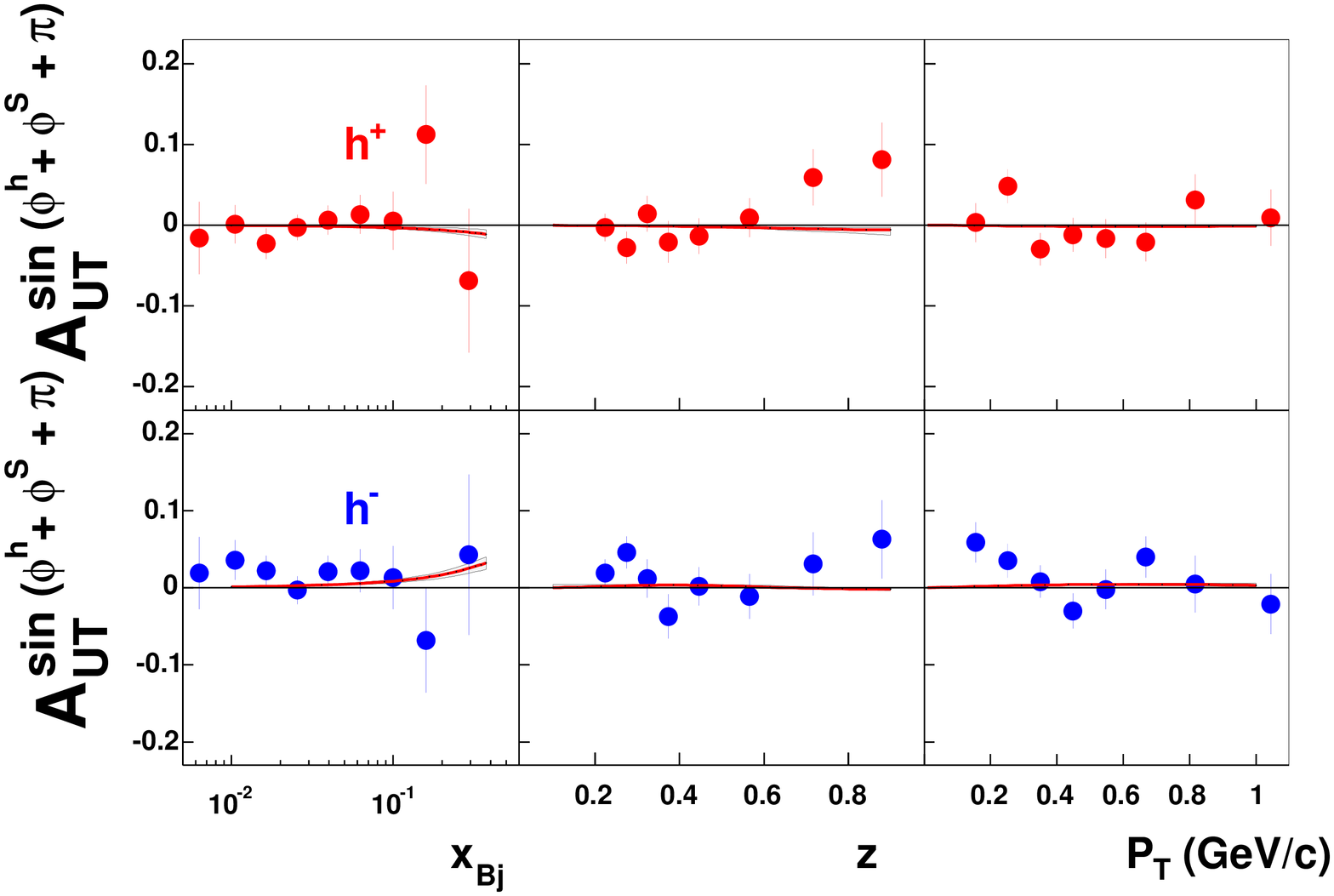} 
\caption{\label{fig:hermes_compass} 
Description of HERMES \cite{Airapetian:2004tw}, left, and COMPASS \cite{Alexakhin:2005iw}, right, 
data.} 
\end{figure}
\begin{figure}
\includegraphics[width=0.35\textwidth,height=0.45\textwidth,bb= 10 140 540 660,angle=-90]
{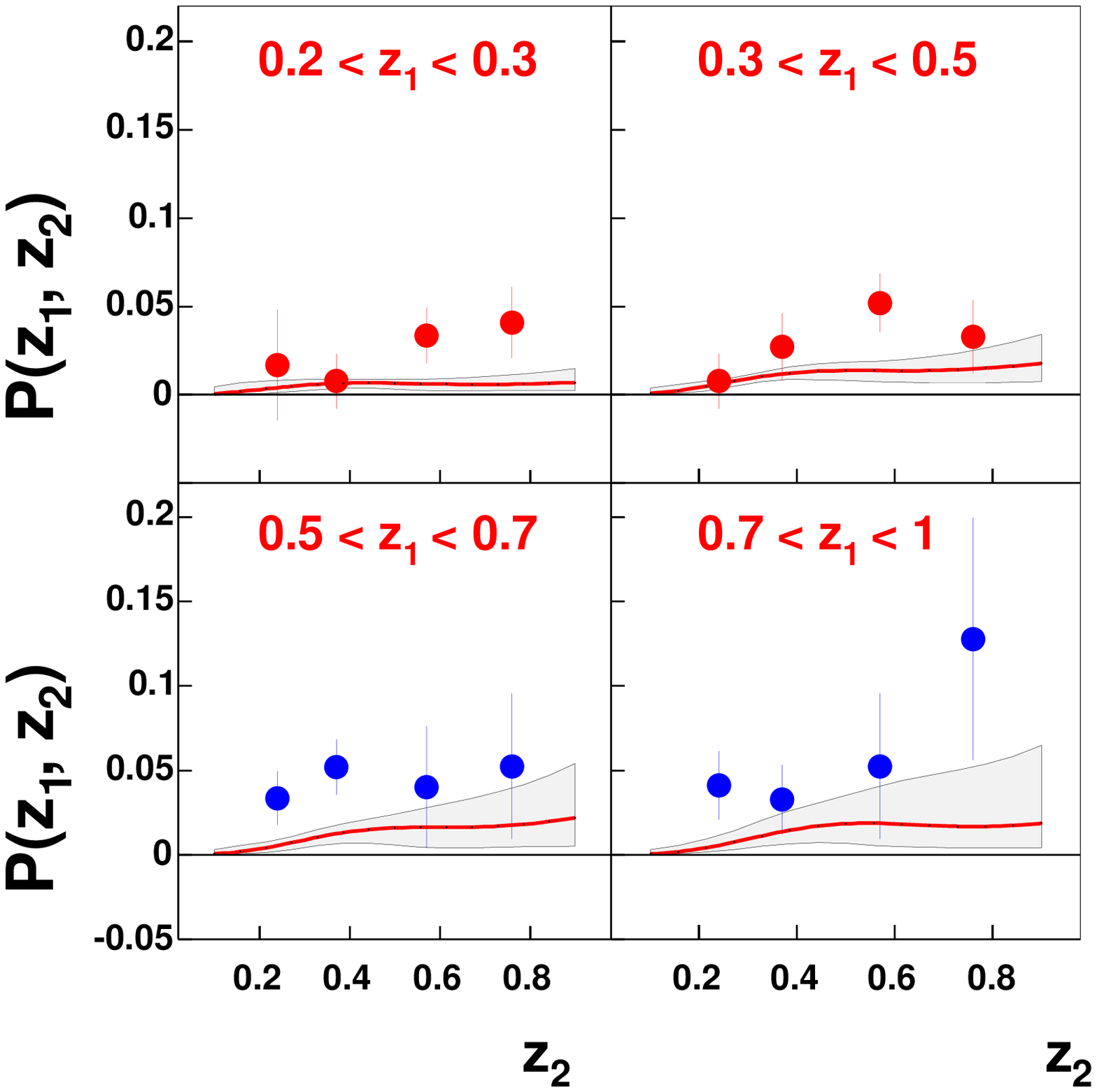} \hskip 0.6cm
\includegraphics[width=0.35\textwidth,height=0.4\textwidth,bb= 10 140 540 660,angle=-90]
{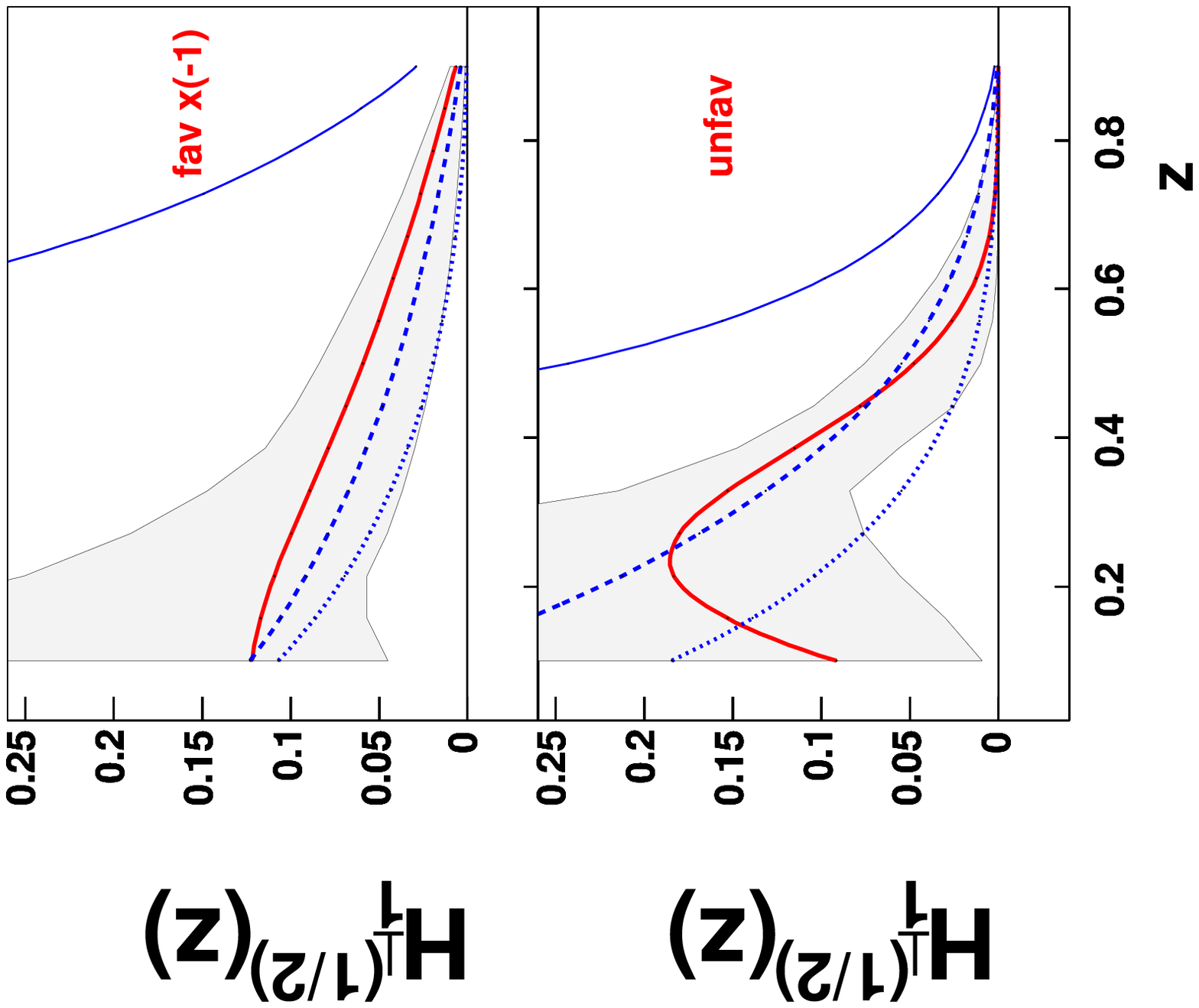} 
\caption{\label{fig:transversity} 
Description of BELLE, left, \cite{Abe:2005zx} data and $1/2$-moment of the Collins fragmentation function compared to results of Refs. \cite{Vogelsang:2005cs} (dotted line) and \cite{Efremov:2006qm} (dashed line).} 
\end{figure}

The values of the parameters as determined through our fit are shown in 
Table \ref{tab:a}.
The results of the fit of HERMES data are presented in the left panel of Fig. \ref{fig:hermes_compass} and the predictions for COMPASS are compared to experimental data in the right panel of Fig. \ref{fig:hermes_compass}. BELLE data on 
$e^+e^-\to h_1 h_2 X$ azimuthal asymmetry are compared to the
model calculations in Fig. \ref{fig:transversity}. The data are slightly underestimated indicating that the transversity distribution does not saturate the Soffer bound. Results on the extraction of the transversity distribution from experimental data will be published elsewhere, Ref. \cite{our}.

\begin{table}[t]
\begin{tabular}{c|l|l|l|l}
\hline
$\Delta^N D $ & $N_{fav}=$ -0.66 $\pm$ 0.52    & 
$\alpha_{fav}=$ 1.24 $\pm$ 0.9 & $\beta_{fav}=$ 0  $\pm$     4  & 
\\ \hline
  &  $N_{unf}=$ 0.52  $\pm$     0.15   & 
$\alpha_{unf}=$ 3.2 $\pm$ 3.4 & $\beta_{unf}=$ 4.1 $\pm$ 5.5  & 
\\ \hline
(GeV$^2$) & $M_h^2=$ 0.56 $\pm$ 0.15 & 
$\langle\kt{^2}\rangle\T=$ 0.28 & 
$\langle\kt{^2}\rangle=$  0.28 \tablenote{See Ref. \cite{Anselmino:2006rv}} & 
$\langle p_\perp{^2}\rangle=$  0.25$^*$ 
\\ \hline
\end{tabular}
\caption{Set of free parameters as determined by the fit.}
\label{tab:a}
\end{table}

\begin{figure}
\includegraphics[width=0.35\textwidth,bb= 10 140 540 660,angle=-90]
{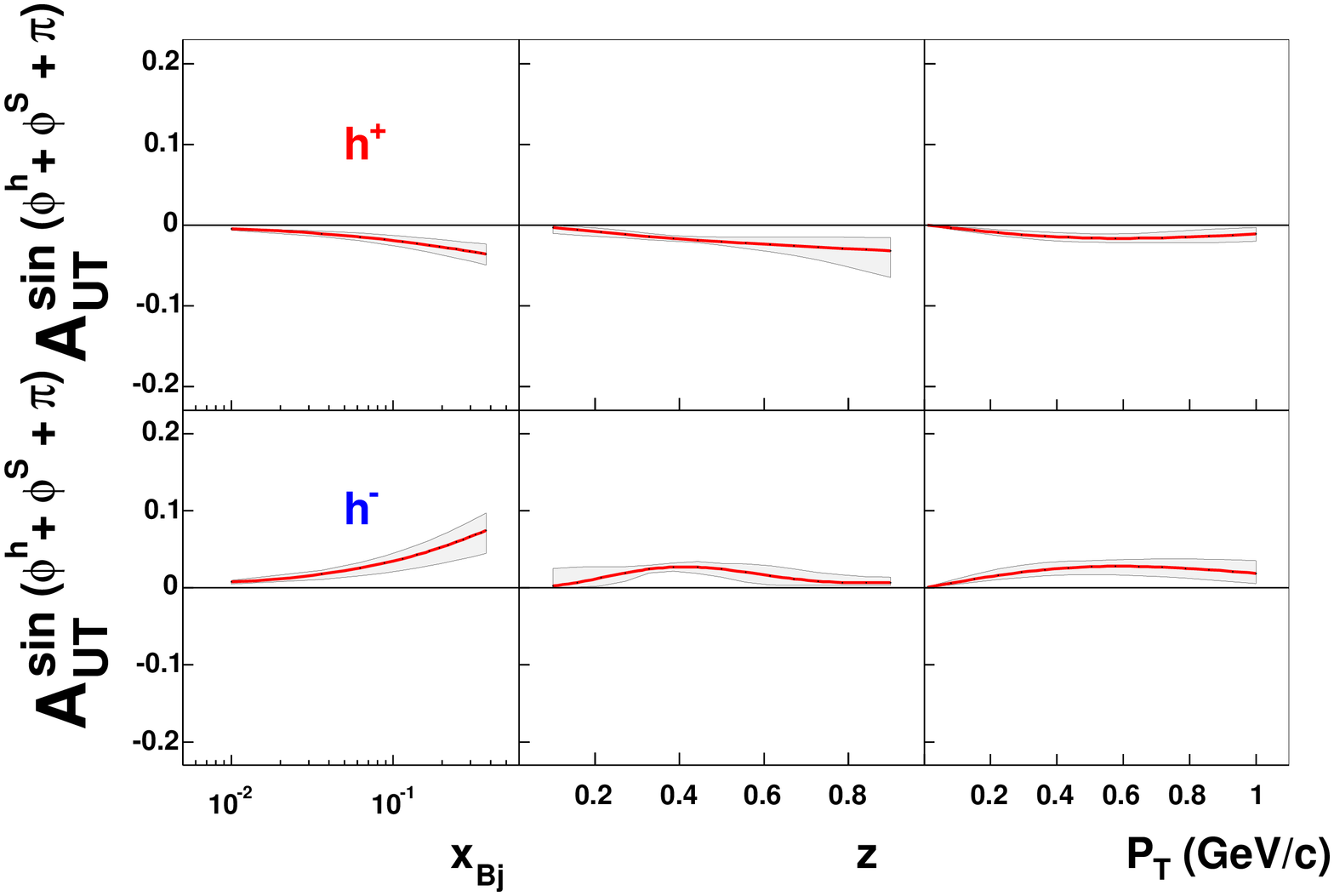} \hskip 1.85cm
\includegraphics[width=0.35\textwidth,bb= 10 140 540 660,angle=-90]
{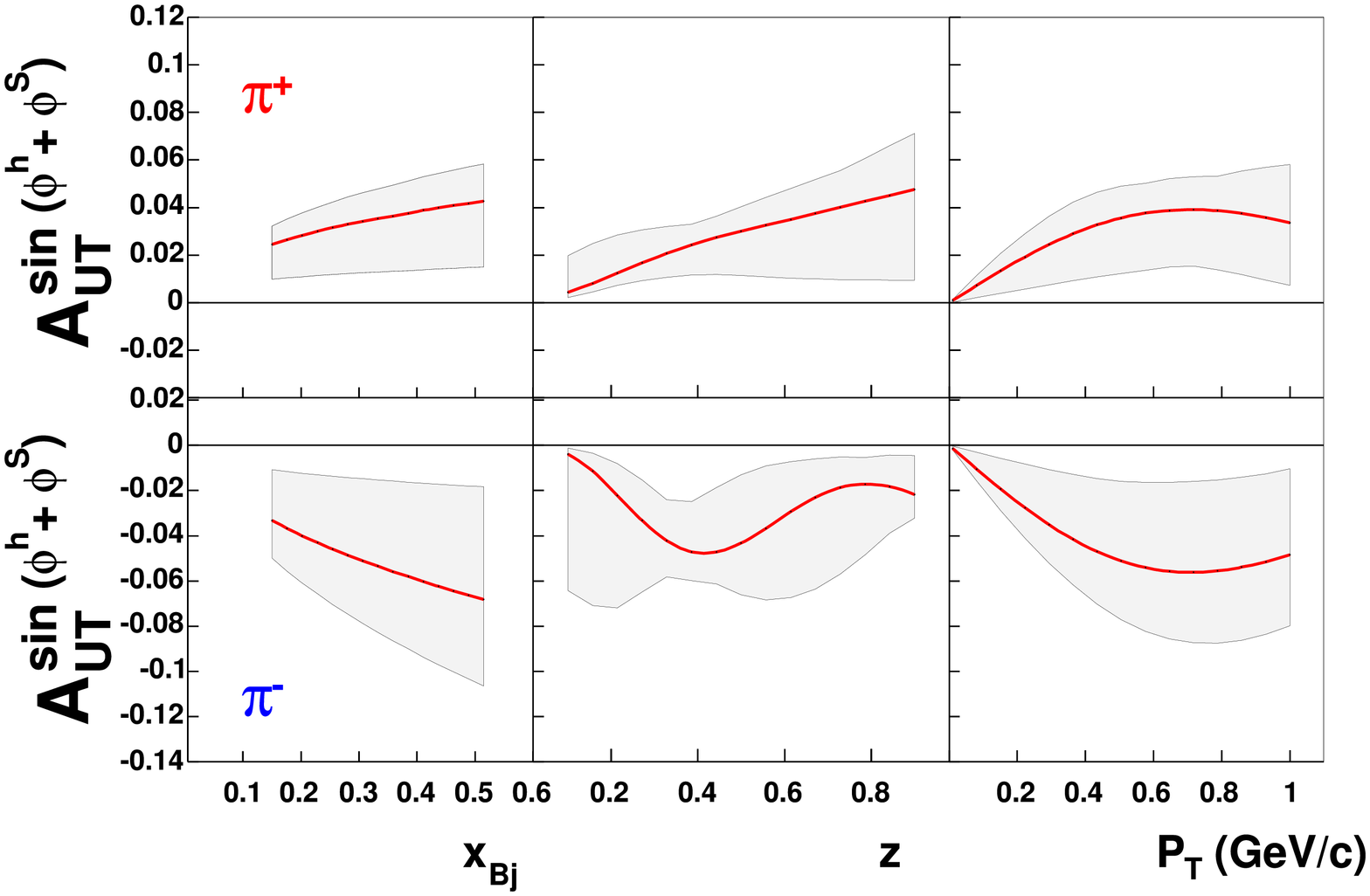} 
\caption{\label{fig:compass_jlab} 
Predictions for COMPASS, left, operating with proton target and JLab, right, at 6 GeV.} 
\end{figure}

In 
Fig. \ref{fig:compass_jlab} we show our predictions for COMPASS experiment operating with a proton target and JLAB at 6 GeV.
 
\paragraph{\bf Conclusions}

Using the available data on Collins effect in polarized SIDIS it is possible to extract 
the Collins fragmentation functions.  
Our results on the Collins fragmentation functions are in agreement with those  of other authors \cite{Vogelsang:2005cs,Efremov:2006qm}.
The extracted functions are compatible with BELLE \cite{Abe:2005zx} data for $e^+e^-\to h_1 h_2 X$. New precise measurements of the Collins asymmetry \cite{Ageev:2006da,Seidl} allow to extract the transversity distribution, Ref. \cite{our}. Sizable asymmetries are expected at JLAB and COMPASS operating with proton target.


\bibliographystyle{aipproc}   


\begin{thebibliography}{17}
\expandafter\ifx\csname natexlab\endcsname\relax\def\natexlab#1{#1}\fi
\providecommand{\enquote}[1]{``#1''}
\expandafter\ifx\csname url\endcsname\relax
  \def\url#1{\texttt{#1}}\fi
\expandafter\ifx\csname urlprefix\endcsname\relax\def\urlprefix{URL }\fi
\providecommand{\eprint}[2][]{\url{#2}}

\bibitem[Airapetian et~al.(2005)]{Airapetian:2004tw}
A.~Airapetian, et~al., \emph{Phys. Rev. Lett.} \textbf{94}, 012002 (2005).

\bibitem[Alexakhin et~al.(2005)]{Alexakhin:2005iw}
V.~Y. Alexakhin, et~al., \emph{Phys. Rev. Lett.} \textbf{94}, 202002 (2005),
  \eprint{hep-ex/0503002}.

\bibitem[Collins(1993)]{Collins:1992kk}
J.~C. Collins, \emph{Nucl. Phys.} \textbf{B396}, 161--182 (1993).

\bibitem[Bacchetta et~al.(2004)]{Bacchetta:2004jz}
A.~Bacchetta, U.~D'Alesio, M.~Diehl, and C.~A. Miller, \emph{Phys. Rev.}
  \textbf{D70}, 117504 (2004), \eprint{hep-ph/0410050}.

\bibitem[Abe et~al.(2006)]{Abe:2005zx}
K.~Abe, et~al., \emph{Phys. Rev. Lett.} \textbf{96}, 232002 (2006).

\bibitem[Kotzinian(1995)]{Kotzinian:1994dv}
A.~Kotzinian, \emph{Nucl. Phys.} \textbf{B441}, 234--248 (1995),
  \eprint{hep-ph/9412283}.

\bibitem[Mulders and Tangerman(1996)]{Mulders:1995dh}
P.~J. Mulders, and R.~D. Tangerman, \emph{Nucl. Phys.} \textbf{B461}, 197--237
  (1996).

\bibitem[Boer et~al.(1997)]{Boer:1997mf}
D.~Boer, R.~Jakob, and P.~J. Mulders, \emph{Nucl. Phys.} \textbf{B504},
  345--380 (1997).

\bibitem[Soffer(1995)]{Soffer:1994ww}
J.~Soffer, \emph{Phys. Rev. Lett.} \textbf{74}, 1292--1294 (1995).

\bibitem[Gluck et~al.(1998)]{Gluck:1998xa}
M.~Gluck, E.~Reya, and A.~Vogt, \emph{Eur. Phys. J.} \textbf{C5}, 461--470
  (1998).

\bibitem[Gluck et~al.(2001)]{Gluck:2000dy}
M.~Gluck, E.~Reya, M.~Stratmann, and W.~Vogelsang, \emph{Phys. Rev.}
  \textbf{D63}, 094005 (2001).

\bibitem[Anselmino et~al.(2006{\natexlab{a}})]{our}
M.~Anselmino, M.~Boglione, U.~D'Alesio, A.~Kotzinian, F.~Murgia, A.~Prokudin,
  and C.~Turk, \emph{in preparation}.

\bibitem[Anselmino et~al.(2006{\natexlab{b}})]{Anselmino:2006rv}
M.~Anselmino, M.~Boglione, A.~Prokudin, and C.~Turk,
  \eprint{hep-ph/0606286}.

\bibitem[Vogelsang and Yuan(2005)]{Vogelsang:2005cs}
W.~Vogelsang, and F.~Yuan, \emph{Phys. Rev.} \textbf{D72}, 054028 (2005).

\bibitem[Efremov et~al.(2006)]{Efremov:2006qm}
A.~V. Efremov, K.~Goeke, and P.~Schweitzer, \emph{in these proceedings {\rm
  and} Phys. Rev.} \textbf{D73}, 094025 (2006).

\bibitem[Ageev et~al.(2006)]{Ageev:2006da}
E.~S. Ageev, et~al., \eprint{hep-ex/0610068}.

\bibitem[Seidl(2006)]{Seidl}
R.~Seidl, \emph{in these proceedings}.

\end{thebibliography}

\end{document}